\documentclass[12pt,article,nofootinbib]{revtex4}
\usepackage{pdfpages}
\usepackage{epsf}
\usepackage{subfigure}
\usepackage{amsmath}
\usepackage{eqnarray}
\usepackage{epstopdf}
\usepackage{mathrsfs}
\usepackage{natbib}
\usepackage{amssymb}
\usepackage{dcolumn}
\usepackage{graphicx}
\usepackage{color} 
\usepackage{hyperref}
\usepackage{enumerate}

\makeatletter
\newcommand*{\rom}[1]{\expandafter\@slowromancap\romannumeral #1@}
\makeatother
\def\be{\begin{equation}}
\def\ee{\end{equation}}
\def\ba{\begin{eqnarray}}
\def\ea{\end{eqnarray}}

\graphicspath{{images/}}
\begin{document}

	\title{\large \bf The Interpretability of LSTM Models for Predicting Oil Company Stocks: Impact of Correlated Features}

	\author{Javad T. Firouzjaee}
	\affiliation{Department of Physics, K. N. Toosi University of Technology, P. O. Box 15875-4416, Tehran, Iran}
	\affiliation{PDAT Laboratory, Department of Physics, K. N. Toosi University of Technology, P. O. Box 15875-4416, Tehran, Iran}
	\affiliation{ School of Physics, Institute for Research in Fundamental Sciences (IPM), P. O. Box 19395-5531, Tehran, Iran } 
		\email{firouzjaee@kntu.ac.ir} 
	
	\author{Pouriya Khalilian}
	\affiliation{Department of Physics, K. N. Toosi University of Technology, P. O. Box 15875-4416, Tehran, Iran}
	\affiliation{PDAT Laboratory, Department of Physics, K. N. Toosi University of Technology, P. O. Box 15875-4416, Tehran, Iran}	
		\email{pouriya@email.kntu.ac.ir}

\begin{abstract}
Oil companies are among the largest companies in the world whose economic indicators in the global stock market have a great impact on the world economy\cite{ec00} and market due to their relation to gold\cite{ec01}, crude oil\cite{ec02}, and the dollar\cite{ec03}. This study investigates the impact of correlated features on the interpretability of Long Short-Term Memory(LSTM)\cite{ec04} models for predicting oil company stocks. To achieve this, we designed a Standard Long Short-Term Memory (LSTM) network and trained it using various correlated datasets. Our approach aims to improve the accuracy of stock price prediction by considering the multiple factors affecting the market, such as crude oil prices, gold prices, and the US dollar. The results demonstrate that adding a feature correlated with oil stocks does not improve the interpretability of LSTM models. These findings suggest that while LSTM models may be effective in predicting stock prices, their interpretability may be limited. Caution should be exercised when relying solely on LSTM models for stock price prediction as their lack of interpretability may make it difficult to fully understand the underlying factors driving stock price movements. We have employed complexity analysis to support our argument, considering that financial markets encompass a form of physical complex system\cite{ec05}. One of the fundamental challenges faced in utilizing LSTM models for financial markets lies in interpreting the unexpected feedback dynamics within them.
 
\end{abstract}
\maketitle

\newpage
\section{Introduction}

Machine Learning, which is one of the subfields of Artificial Intelligence, has its applications in various fields including Economics, Medicine \cite{ec1}, Cosmology \cite{ec2}, Photonics \cite{ec3}, Robotics \cite{ec4},IoT(Internet of Things)\cite{Vochozka0} and etc. Long Short-Term Memory (LSTM) models are widely used in the financial industry for time series prediction, including stock prices. LSTMs are known for their ability to capture complex patterns in sequential data. However, one major limitation of LSTM models is their lack of interpretability. That is, it is often difficult to understand how the model arrives at its predictions and what factors are driving the results. Despite their lack of interpretability, the performance of LSTMs in predicting stock prices has been well-documented in previous research \cite{ecq1},\cite{ecq2}. Therefore, while caution is necessary when relying solely on LSTMs for financial decisions, these models can still be useful in predicting the future price movements of stocks. 

It is widely acknowledged that oil companies occupy the largest market share in the energy industry\cite{ec010}. These companies have enormous global reach and wield substantial influence in the world economy. In fact, oil is arguably the most vital factor of the global economy, as the imposition of sanctions on the export or import of oil from many countries can practically paralyze their economies, especially for those countries with oil-dependent economies, such as the Persian Gulf countries\cite{ec06}. Additionally, stock indices of oil companies are regarded as among the most important indicators in the global stock market, with correlations between the shares of oil companies, gold, the US dollar, and crude oil, all of which have a significant impact on the market\cite{ec07}.

Neural Networks, despite their lack of interpretability, have proven useful in modeling oil companies. Long Short-Term Memory(LSTM) neural networks are one of the most powerful architectures in Recurrent Neural Networks. They have addressed the problem of gradient vanishing in recurrent neural networks\cite{ec08}, thereby allowing more accurate predictions of the stock market\cite{ec09}. This market behaves like an unstable dynamic system, characterized by numerous non-linear correlations. Accordingly, researchers have proposed various methods to predict its behavior.

To review the literature oil research path, we can mention Alvarez-Ramirez et al. work \cite{Alvarez}  which analyzed the auto-correlations of international crude oil.
After carrying out several tests, in 2005, Moshiri and Foroutan \cite{ec5} concluded that oil stock markets have a recursive architecture because they are time series. They used three methods ANN, GARCH, and ARMA, and the best results come from the ANN method.

Author in \cite{ec6} published an article using Multilayer Neural Networks, which examined the relationship between crude oil prices and current prices.
Moreover, they showed that the future prices of crude oil contain new information about oil spot price detection. 
Ye et al. \cite{Ye} studied the changing relationships between the prices of crude oil and several other factors from January 1992 to December 2007 by the Short-Run Crude Oil Price Forecast Model.

Chen and colleagues developed a model based on deep learning and used this model to model the unknown non-linear behavior of WTI stocks \cite{ec8}.
QI, KHUSHI, and  POON used different recursive neural network architectures including LSTM, GRU, BiLSTM, RNN to model the Forex market and obtained significant results from these models to predict several currency pairs. They used a database that relates to ELLIOT method information, one of the stock market forecasting methods \cite{ec9}.\\

In 2018, Gupta and Pandey predicted crude oil prices by using LSTM network \cite{Gupta}, and following that Cen and Wang applied deep learning algorithms to anticipate the volatility behavior of crude oil prices \cite{Cen}. To solve the chaotic and nonlinear features of crude oil time series Altan and Karasu used a new crude oil price prediction model is proposed in this study, which includes the long short-term memory (LSTM), technical indicators \cite{Altan}.
\\

In 2017, Arfaoui and  Rejeb published an article examining the effects and relationships of stock markets, oil, dollars, and gold based on global market documentation.
They concluded that oil prices are significantly affected by stock markets, gold, and the dollar and that there are always indirect effects, which also confirms the presence of correlations in the global market\cite{Aymen}. In this paper, we compare this correlation feature and the relationships between stocks with the dollar, crude oil, gold, and major oil company stock indices, we create datasets and compare the results of forecasts with real data.
\\

This \cite{smith} article dives deep into the challenges posed by LSTM models in financial market forecasting. Smith et al. discuss the concept of "black box" models - LSTM architectures that lack transparency in their decision-making process. By examining the weights, biases, and internal mechanisms of LSTMs, the authors ascertain the difficulties in understanding and interpreting these models, especially in financial contexts. They emphasize the need for interpretability, highlighting potential risks and the consequences of solely relying on uninterpretable LSTM models.
\\

Johnson, et al. tackle the enigma of LSTM interpretability within the financial markets domain. They discuss the challenges of interpreting LSTM models specifically for financial forecasting. The authors emphasize the importance of explaining predictions and making sense of underlying patterns and features to build trust in financial decision-making. The article provides a holistic overview of conceptual and technical difficulties, articulating the limitations of LSTM models in this context\cite{johnson}.
\\

Brown et al. delve into the intricate aspect of feature importance within LSTM models. They argue that understanding which features are most influential in driving predictions is vital for financial market forecasting. The article explores various techniques and approaches that attempt to unravel the complexity of feature transformations within LSTMs. By illuminating the challenges of achieving feature importance analysis, the authors call for further research to address the uninterpretability of LSTM models\cite{brown}.
\\

LSTM is a dynamic network that is used to predict these markets due to the dynamic nature of energy financial markets.The transparency of this algorithm in detecting suprising based on stochastic distortion\cite{ec011} is a fundamental challenge for investors.Considering the importance of oil company stocks for investors, checking the transparency of this algorithm can be significant.Considering that there are various causes for fluctuations and disruptions in the markets of oil companies, the investigation of the main economic factors in the comparative impact of these stocks can be of great help in modeling.The purpose of this article is to investigate the effect of these main economic factors that have correlation with the stocks of major oil companies and to investigate the effect on the interpretability for regression with LSTM.These articles collectively establish the significance of interpretability in LSTM models for financial market forecasting. They consistently highlight the limitations and potential risks of relying solely on uninterpretable LSTM models and advocate for the need to strike a balance between accuracy and interpretability.
\\

This article is structured as follows: First, we investigate the complexity of the relationships between stocks through discrete correlation analysis. Subsequently, we design a standard LSTM network, optimize its hyper-parameters, and then perform modeling. Finally, we review the analysis and modeling results and draw conclusions.

\section{Oil Shares and complexity of stocks relations}
The relation between oil companies' stocks on one hand and the prices of other assets like gold, crude oil, and the price of the US dollar, on the other hand, is intrinsically interesting for many economical and political reasons.
Before calculating the complexity, it has worth having a short introduction for oil companies. 

\subsection{Oil companies}
 In the global economy, large companies operating in the field of energy and oil are among the most authoritative companies in the global economy. Companies such as Total, BP, Cairn Energy, Schlumberger are among the most authoritative companies. These companies are trading in energy and oil and their investors and shareholders are among the largest investors and shareholders in the world, for whom the trend of changing the indicators of these companies in the stock market is very worthful. Moreover, the strong dependence of industry\cite{Vochozka} in various fields on oil has made oil a major factor influencing the policies of each country. In the following, we introduce some of these companies which we analyze their shares in this paper.\\

WTI:  One of known light sweet \textit{crude oil} is West Texas Intermediate (WTI) which refers as one of the main global oil benchmarks. WTI has hight quality is  easy to refine and is sourced primarily from inland Texas.\\

BP.L: BP plc is a British multinational oil and gas company based in London, England. It is one of the world's seven oil and gas companies.\\

FP.PA: TotalEnergies SE is a French multinational unified oil and gas company which is established in 1924 and is one of the seven major oil companies.\\

SLB.PA: Schlumberger Limited is an oil field services company whose member have more than 140 nationalities working in more than 120 countries. Schlumberger executive offices are located in Paris, Houston, London, and The Hague.\\

\subsection{Correlation between shares and economic stocks}\label{corr}

The crude oil index is the most important characteristic in determining the index of oil companies. The sale of crude oil in the world market is done in dollars. Commonly, the dollar can change the value of the indexes of oil companies. Moreover, the gold value can cause changes and fluctuations in currency indices. Their effects on each others can be obtained by calculating the correlation coefficient between their data \cite{Aymen}. In this article, the value of the stock indices that we have examined, from 08/03/2009 to 07/01/2021, the results of the correlation calculation are shown in the table \eqref{corr-table}:

\begin{table}[!htb]
	\caption{Correlations between Shares.}
	\label{corr-table}
	\begin{tabular} {l|l|l|l|l|l|l|l}
		& USD &	WTI &	GOLD &	BP &	TOTAL&	Schlumberger &Cairn Energy   \\ \hline
		USD&	1.000000&	-0.357051&	-0.002025&	-0.214303&	-0.088397&	0.225169&	0.046451\\
		WTI&	-0.357051&	1.000000&	-0.046961&	0.371175&	0.355021&	0.377098&	0.490978\\
		GOLD&-0.002025&	-0.046961&	1.000000&	-0.464169&	-0.553079&	-0.514969&	-0.141593\\
		BP	&-0.214303	&0.371175	&-0.464169	&1.000000	&0.871033	&0.381587	&0.521461\\
		TOTAL&	-0.088397&	0.355021&	-0.553079&	0.871033&	1.000000&	0.529159&	0.494148\\
		Schlumberger&	0.225169&	0.377098&	-0.514969&	0.381587&	0.529159&	1.000000&	0.431568	\\
		Cairn Energy&	0.046451&	0.490978&	-0.141593&	0.521461&	0.494148&	0.431568&	1.000000	\\

	\end{tabular}
\end{table}

This table is represented as an array where its principal diameter is equal to one, and the other component represents the correlation coefficient between the value of stock indices introduced in the global market.\\

\begin{figure}[!htb]
	\centering
	\includegraphics[width=0.8 \columnwidth]{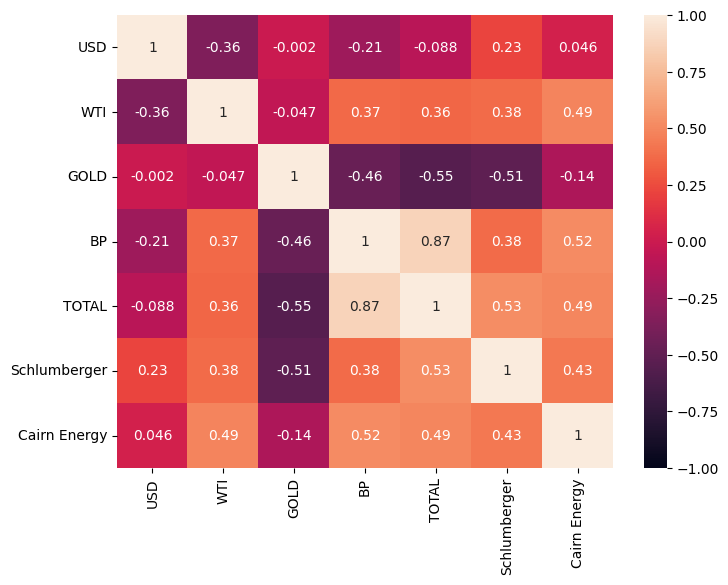}
	\caption{Heatmap of the correlation coefficients in the Table \eqref{corr-table}.}
	\label{corr-heatmap}
\end{figure}

To have a better presentation, we draw the \textbf{Heatmap} diagram which is a type of information visualization (data visualization) in which the value of each cell of the matrix input is displayed in one color. The correlation coefficient obtained in the previous part is shown in the diagram of the thermal map below Fig. \eqref{corr-heatmap}. The lighter the color, the more direct the relationship and the higher the direct correlation, and the darker color, the higher the inverse relationship and correlation.
\\

\subsection{Discrete correlation analysis for oil shares}\label{dcorr}
To conduct a detailed analysis of the relationships between WTI, USD, and GOLD with shares of TOTAL, BP, Schlumberger, and Cairn Energy oil companies, we discretized the data and then examined the Pearson correlations between them. Between 2013/6/24 and 2021/4/7, we discretized around 2000 data points for each stock into 50 40-day data sets.

\begin{figure}[!htb]
	\centering
	\includegraphics[width=1.10 \columnwidth]{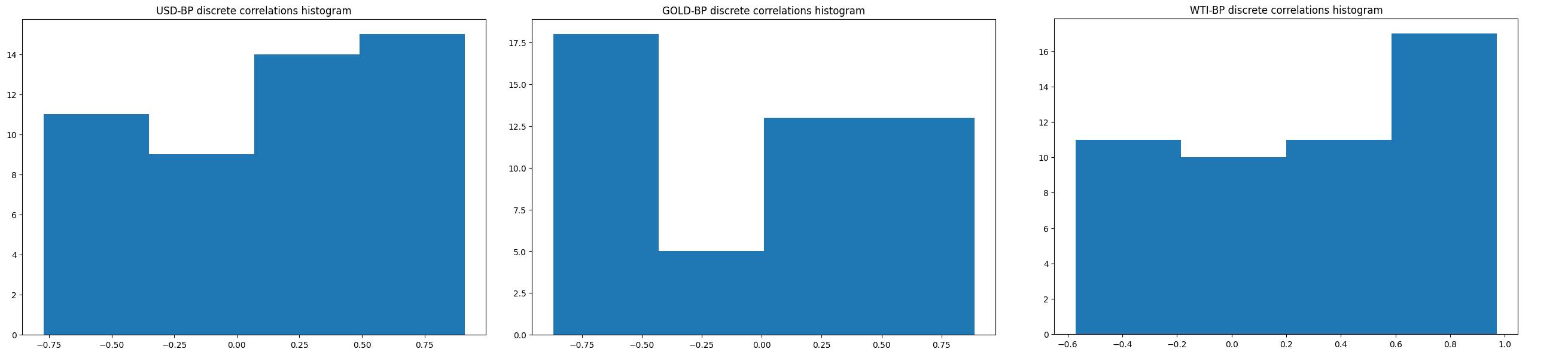}
	\caption{BP discrete correlations histograms .}
	\label{bp-hist}
\end{figure}

\begin{figure}[!htb]
	\centering
	\includegraphics[width=0.8 \columnwidth]{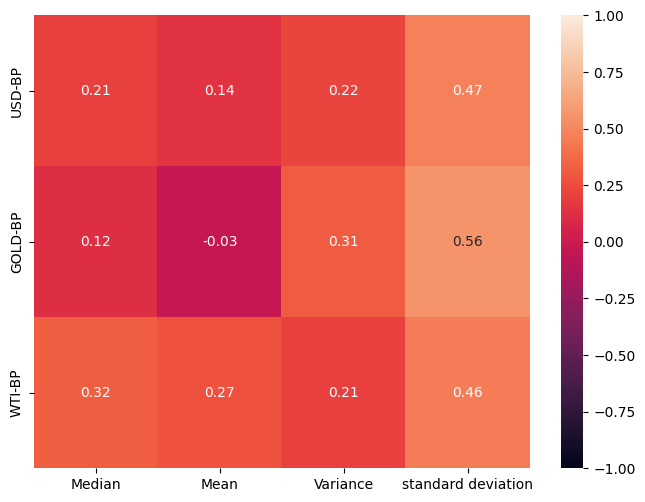}
	\caption{BP discrete correlations statical parameters analysis heatmap .}
	\label{bp-static}
\end{figure}

\begin{figure}[!htb]
	\centering
	\includegraphics[width=1.10 \columnwidth]{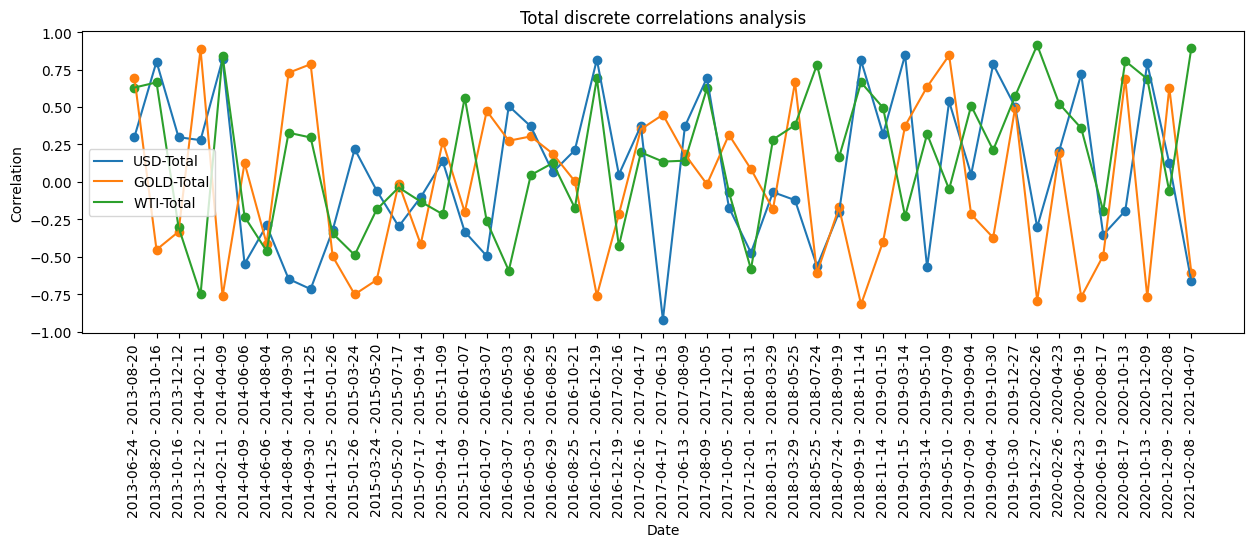}
	\caption{Total discrete correlations analysis .}
	\label{totalcorr}
\end{figure}

\begin{figure}[!htb]
	\centering
	\includegraphics[width=1.10 \columnwidth]{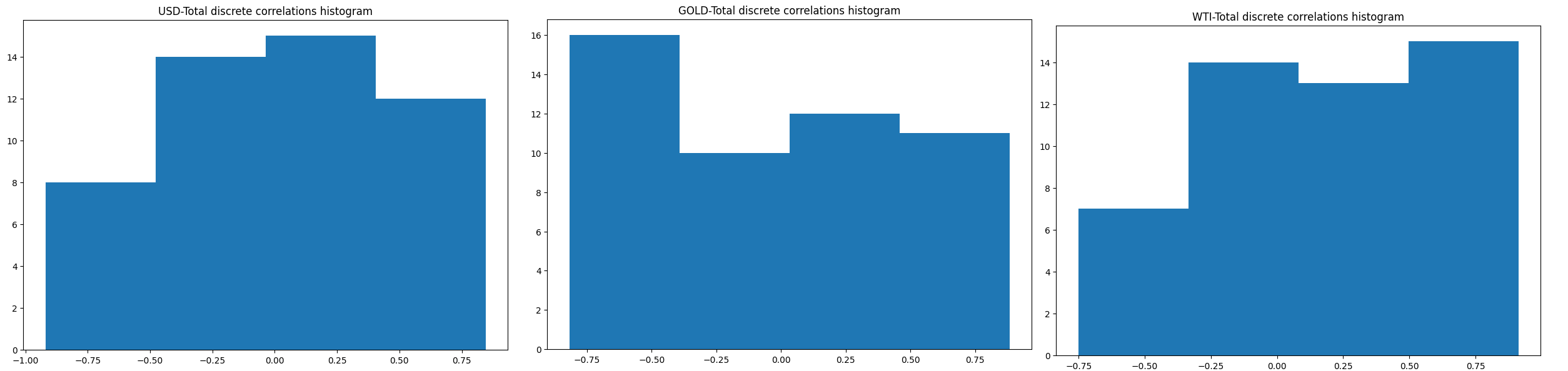}
	\caption{Total discrete correlations histograms .}
	\label{total-hist}
\end{figure}

\begin{figure}[!htb]
	\centering
	\includegraphics[width=0.8 \columnwidth]{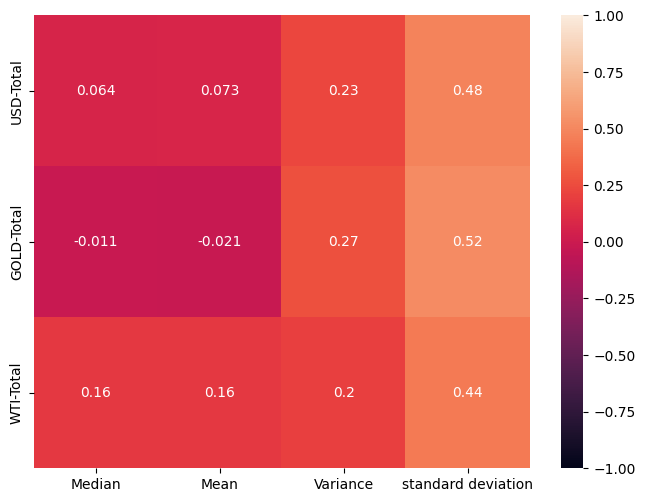}
	\caption{Total discrete correlations statistical parameters analysis heatmap .}
	\label{total-static}
\end{figure}

\begin{figure}[!htb]
	\centering
	\includegraphics[width=1.10 \columnwidth]{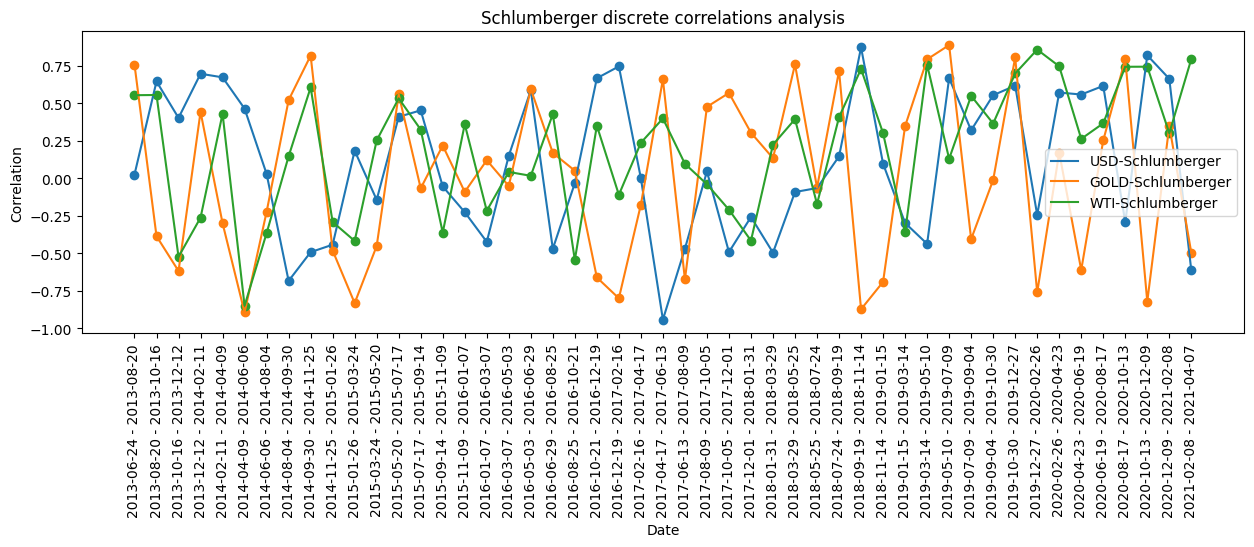}
	\caption{Schlumberger discrete correlations analysis .}
	\label{slbcorr}
\end{figure}

\begin{figure}[!htb]
	\centering
	\includegraphics[width=1.10 \columnwidth]{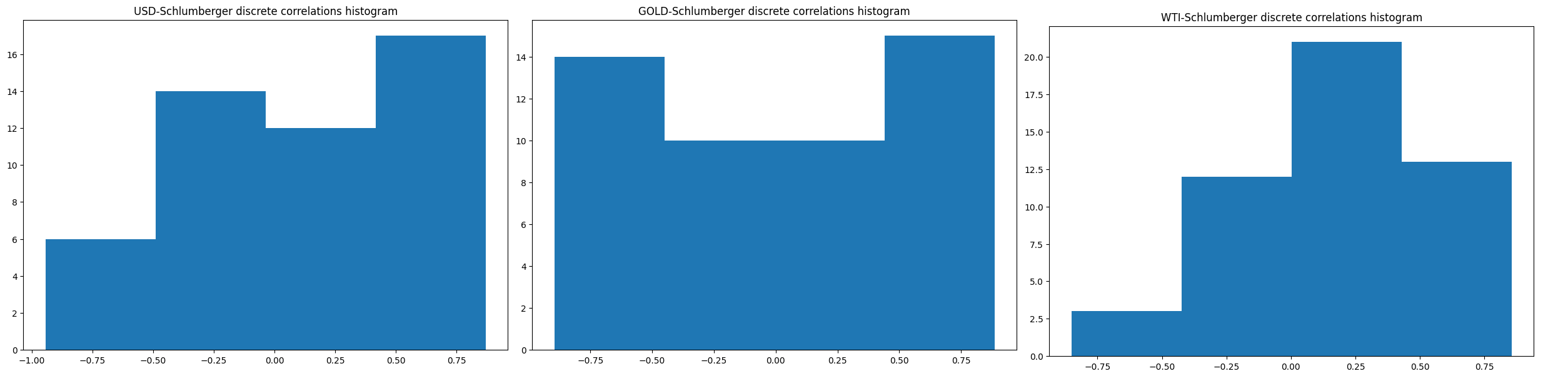}
	\caption{Schlumberger discrete correlations histograms .}
	\label{slb-hist}
\end{figure}

\begin{figure}[!htb]
	\centering
	\includegraphics[width=0.8 \columnwidth]{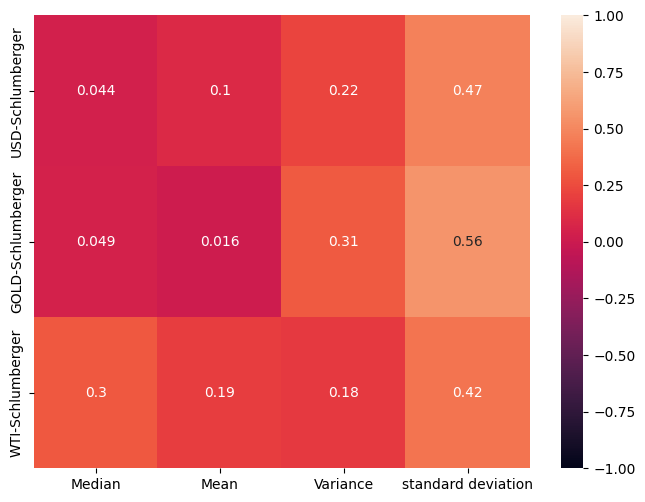}
	\caption{Schlumberger discrete correlations statistical parameters analysis heatmap .}
	\label{slb-static}
\end{figure}

\begin{figure}[!htb]
	\centering
	\includegraphics[width=1.10 \columnwidth]{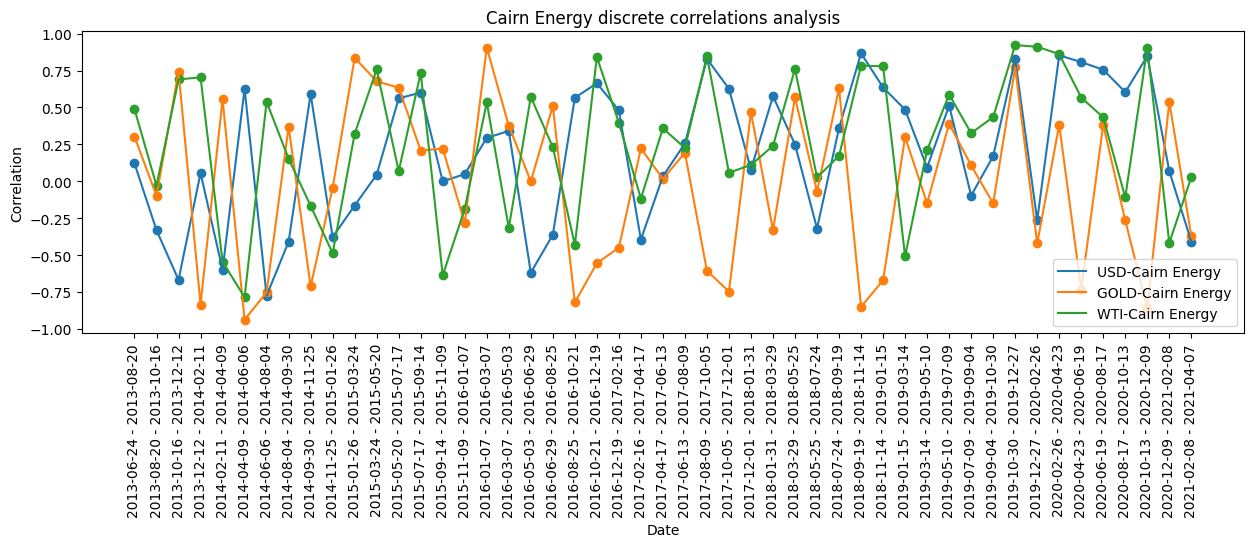}
	\caption{Cairn Energy discrete correlations analysis .}
	\label{cnecorr}
\end{figure}

\begin{figure}[!htb]
	\centering
	\includegraphics[width=1.10 \columnwidth]{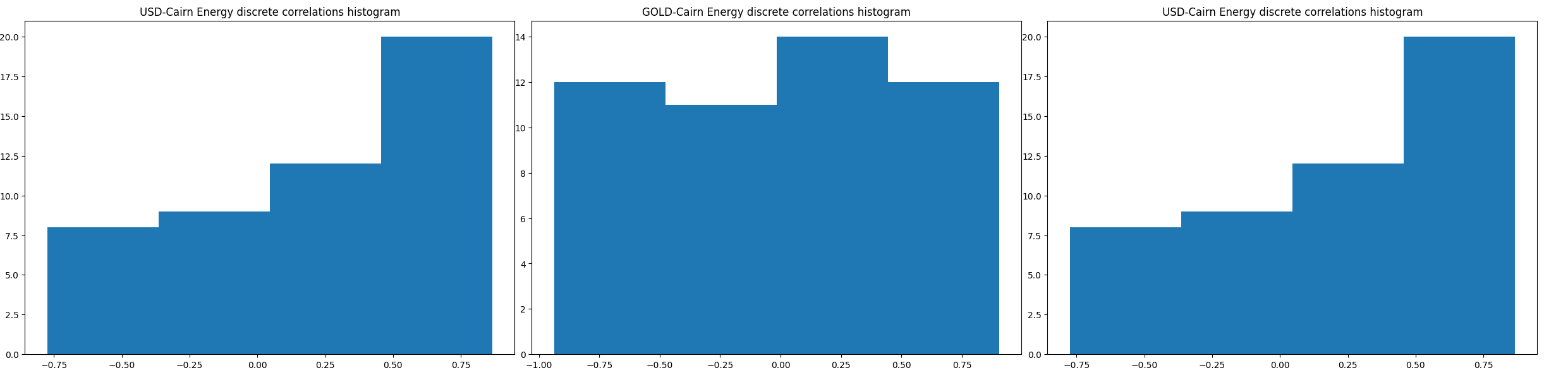}
	\caption{Cairn Energy discrete correlations histograms .}
	\label{cne-hist}
\end{figure}

\begin{figure}[!htb]
	\centering
	\includegraphics[width=0.8 \columnwidth]{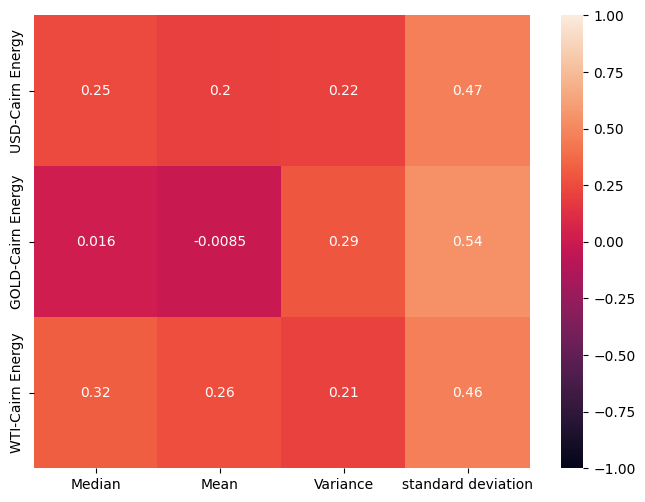}
	\caption{Cairn Energy discrete correlations statistical parameters analysis heatmap .}
	\label{cne-static}
\end{figure}

The FIG\eqref{bpcorr} below presents the discrete correlations of the BP company data set with WTI, USD, and GOLD data sets.\\
\begin{figure}[!htb]
	\centering
	\includegraphics[width=1.10 \columnwidth]{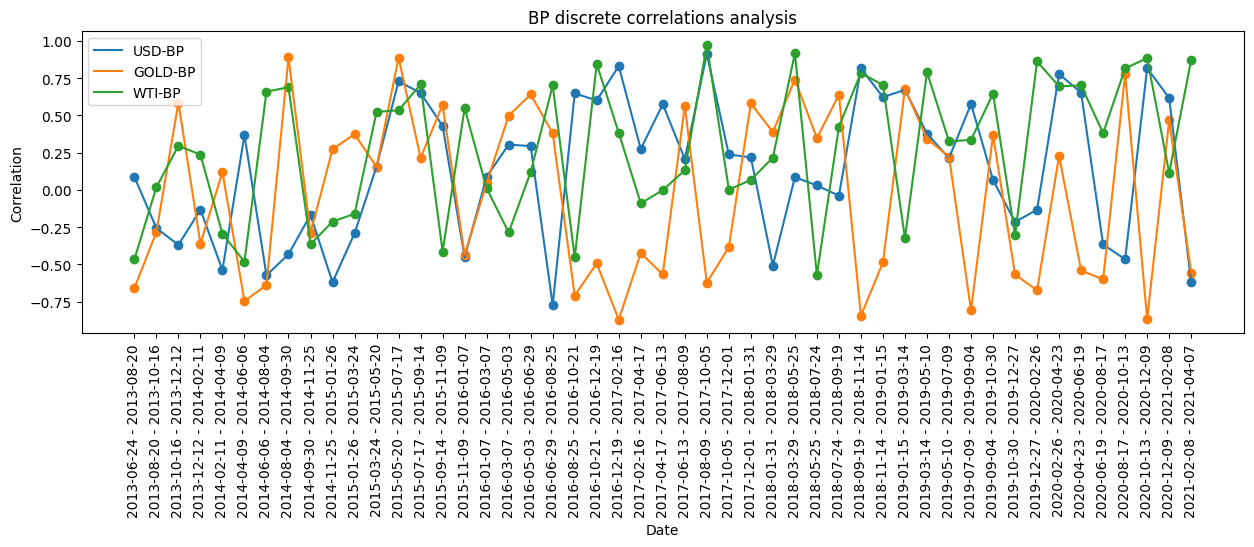}
	\caption{BP discrete correlations analysis .}
	\label{bpcorr}
\end{figure}\\
The correlation histograms of the BP stock index with the WTI, USD, and GOLD stock indices are displayed in FIG\eqref{bp-hist}. The histograms are divided into four columns based on the results of the correlation analysis, as shown in Table\eqref{bp_corr-table}.\\
\\
\begin{table}[!htb]
	\caption{Correlations amount .}
	\label{bp_corr-table}
	\begin{tabular} {l|l|l|l|l}
		&Corr(-1,-0.5)&	Corr(-0.5,0)	&Corr(0,0.5)&	Corr(0.5,1)   \\ \hline
		Count of corr(BP-WTI)&	1.0&	13.0&	15.0&	20.0\\
		Percent of corr(BP-WTI)&	2.0&	26.0&	30.0&	40.0\\
		Count of corr(BP-USD)&	6.0&	12.0&	16.0&	15.0\\
		Percent of corr(BP-USD)&	12.0&	24.0&	32.0&	30.0\\
		Count of corr(BP-GOLD)&	15.0&	8.0&  15.0&	11.0\\
		Percent of corr(BP-GOLD)&	30.0&	16.0&	30.0&	22.0	\\	
	\end{tabular}
\end{table}\\
The statistical parameters, including the average, median, variance, and standard deviation, for the correlations between the BP stock index and the WTI, USD, and GOLD are presented in Table\eqref{bp-static-t} and illustrated in Fig\eqref{bp-static}.
\begin{table}[!htb]
	\caption{BP discrete correlations statical parameters .}
	\label{bp-static-t}
	\begin{tabular} {l|l|l|l|l}
		&Median	&Mean&	Variance&	Standard deviation   \\ \hline
		USD-BP	&0.207317	&0.142085	&0.221153	&0.470269\\
		GOLD-BP	&0.119667	&-0.029858	&0.312736	&0.559228\\
		WTI-BP	&0.325000	&0.274804	&0.209705	&0.457936\\
	
	\end{tabular}
\end{table}\\
\\
The discrete correlations between the Total company data set and the WTI, USD, and GOLD data sets are displayed in FIG.\eqref{totalcorr}.\\
\\
The correlation histograms of the Total stock index with the WTI, USD, and GOLD stock indices are presented in FIG\eqref{total-hist}. The histograms are divided into four columns based on the results of the correlation analysis, as shown in Table\eqref{total_corr-table}.\\
\\
\begin{table}[!htb]
	\caption{Correlations amount of Total .}
	\label{total_corr-table}
	\begin{tabular} {l|l|l|l|l}
		&Corr(-1,-0.5)&	Corr(-0.5,0)	&Corr(0,0.5)&	Corr(0.5,1)   \\ \hline
		Count of corr(Total-WTI)&	3.0&	17.0&	14.0&	15.0\\
		Percent of corr(Total-WTI)&	6.0&	34.0&	30.0&	30.0\\
		Count of corr(Total-USD)&	7.0&	15.0&	15.0&	12.0\\
		Percent of corr(Total-USD)&	14.0&	30.0&	30.0&	26.0\\
		Count of corr(Total-GOLD)&	10.0&	15.0&	15.0&	9.0\\
		Percent of corr(Total-GOLD)&	22.0&	30.0&	30.0&	18.0\\	
	\end{tabular}
\end{table}\\
The statistical parameters, such as the average, median, variance, and standard deviation, for the correlations between the Total stock index and the WTI, USD, and GOLD are displayed in Table\eqref{total-static-t} and Fig\eqref{total-static}.\\
\begin{table}[!htb]
	\caption{Total discrete correlations statical parameters .}
	\label{total-static-t}
	\begin{tabular} {l|l|l|l|l}
		&Median	&Mean&	Variance&	Standard deviation   \\ \hline
	USD-Total&	0.064403&	0.073098&	0.233880&	0.483611\\
	GOLD-Total&	-0.011213&	-0.021289&	0.270061&	0.519674\\
	WTI-Total&	0.164420&	0.164086&	0.195751&	0.442437\\	
	\end{tabular}
\end{table}\\
\\
The discrete correlations between the Schlumberger company data set and the WTI, USD, and GOLD data sets are presented in FIG.\eqref{slbcorr}.
\\
The correlation histograms of the Schlumberger stock index with the WTI, USD, and GOLD stock indices are displayed in FIG\eqref{slb-hist}. The histograms are categorized into four columns based on the results of the correlation analysis, as presented in Table\eqref{slb_corr-table}.\\
\\
\begin{table}[!htb]
	\caption{Correlations amount of Schlumberger .}
	\label{slb_corr-table}
	\begin{tabular} {l|l|l|l|l}
		&Corr(-1,-0.5)&	Corr(-0.5,0)	&Corr(0,0.5)&	Corr(0.5,1)   \\ \hline
		Count of corr(Schlumberger-WTI)&	3.0&	12.0&	21.0&	13.0\\
		Percent of corr(Schlumberger-WTI)&	6.0&	24.0&	42.0&	28.0\\
		Count of corr(Schlumberger-USD)&	3.0&	19.0&	12.0&	15.0\\
		Percent of corr(Schlumberger-USD)&	6.0&	38.0&	26.0&	30.0\\
		Count of corr(Schlumberger-GOLD)&	11.0&	13.0&	12.0&	13.0\\
		Percent of corr(Schlumberger-GOLD)&	22.0&	26.0&	26.0&	26.0\\	
	\end{tabular}
\end{table}\\
The statistical parameters, namely the average, median, variance, and standard deviation, for the correlations of the Schlumberger stock index with the WTI, USD, and GOLD are exhibited in Table\eqref{slb-static-t} and Fig\eqref{slb-static}:\\
\begin{table}[!htb]
	\caption{Schlumberger discrete correlations statistical parameters .}
	\label{slb-static-t}
	\begin{tabular} {l|l|l|l|l}
		&Median	&Mean&	Variance&	Standard deviation   \\ \hline
		USD-Schlumberger&	0.043825&	0.100825&	0.224941&	0.474280\\
		GOLD-Schlumberger&	0.049048&	0.015501&	0.309095&	0.555963\\
		WTI-Schlumberger&	0.299134&	0.193702&	0.175441&	0.418857\\	
	\end{tabular}
\end{table}\\
\\
The discrete correlations between the Cairn Energy company data set and the WTI, USD, and GOLD data sets are presented in FIG.\eqref{cnecorr}.\\
\\
The correlation histograms indicating the relationship between the Cairn Energy stock index and the WTI, USD, and GOLD stock indices are illustrated in FIG\eqref{cne-hist}. The histograms are arranged into four columns based on the results of the correlation analysis presented in Table\eqref{cne_corr-table}.\\
\\
\begin{table}[!htb]
	\caption{Correlations amount of Cairn Energy .}
	\label{cne_corr-table}
	\begin{tabular} {l|l|l|l|l}
		&Corr(-1,-0.5)&	Corr(-0.5,0)	&Corr(0,0.5)&	Corr(0.5,1)   \\ \hline
		Count of corr(Cairn Energy-WTI)&	4.0&	9.0&	18.0&	19.0\\
		Percent of corr(Cairn Energy-WTI)&	8.0&	18.0&	36.0&	38.0\\
		Count of corr(Cairn Energy-USD)&	4.0&	11.0&	17.0&	18.0\\
		Percent of corr(Cairn Energy-USD)&	8.0&	22.0&	34.0&	36.0\\
		Count of corr(Cairn Energy-GOLD)&	12.0&	12.0&	14.0&	12.0\\
		Percent of corr(Cairn Energy-GOLD)&	24.0&	24.0&	28.0&	24.0\\	
	\end{tabular}
\end{table}\\
The statistical parameters, namely the average, median, variance, and standard deviation, for the correlations of the Cairn Energy stock index with the WTI, USD, and GOLD are displayed in Table\eqref{cne-static-t} and Fig\eqref{cne-static}:\\
\begin{table}[!htb]
	\caption{Cairn Energy discrete correlations statically parameters .}
	\label{cne-static-t}
	\begin{tabular} {l|l|l|l|l}
		&Median	&Mean&	Variance&	Standard deviation   \\ \hline
		USD-Cairn Energy&	0.248828&	0.198628&	0.217601&	0.46647\\
		GOLD-Cairn Energy&	0.016285&	-0.008477&	0.290874&	0.539328\\
		WTI-Cairn Energy&	0.321247&	0.261865&	0.214425&	0.463061\\	
	\end{tabular}
\end{table}\\
\\
\section{LSTM architecture for oil prices}

\begin{figure}[!htb]
	\centering
	\includegraphics[width=0.7 \columnwidth]{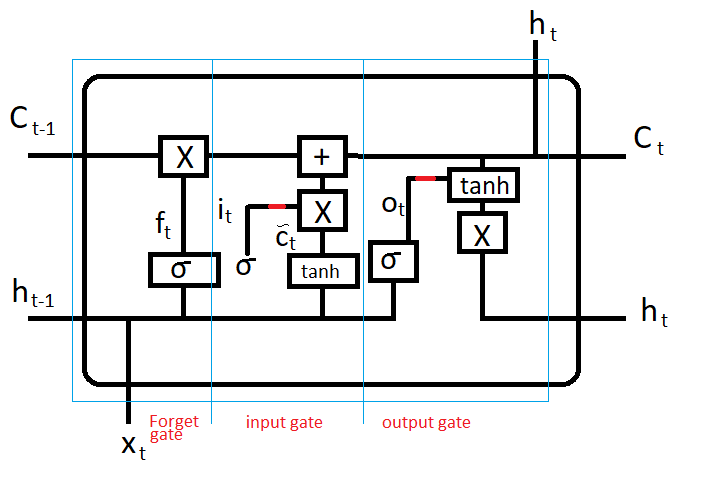}
	\caption{(LSTM Architecture)}
	\label{lstm}
\end{figure}

Long Short Term Memory, LSTM is an architecture for RNNs to avoid the gradient vanishing problem in RNNs in memory building. LSTM architecture has three gates, forget gate, input gate, and output gate as shown in Fig. \eqref{lstm}.  LSTM can read, write and delete information from its memory. The LSTM structure has the capability to identify which cells are stimulated and compressed based on the previous state, available memory, and current input.\\

Forget gatedetermine what information data we're going to throw away from the cell state. In this gate, information is deleted or stored based on the output of the sigmoid function. This gate cell is formulated as follows:

\be
{f_t}: Forget gate, \sigma  : sigmoid ,{W}:Weight,h_{t - 1}:Hidden state
\ee
\be
{x_t}:Input,{b}:Bias ,{i_t}: Input gate ,{c_t}: Cell state,{o_t}:Output gate
\ee

\be
{f_t} = \sigma ({W_f} \times [{h_{t - 1}},{x_t}] + {b _f}).
\ee

Input gate determine what new data we’re  going to store in the cell state. The input gate is actually a gate for writing memory which is represented  as
\be
{i_t} = \sigma ({W_f} \times [{h_{t - 1}},{x_t}] + {b _f}),
\ee

\be
{\tilde c_t} = \tanh ({W_c} \times [{h_{t - 1}},{x_t}] + {b_c}).
\ee
Cell state drops the information about past subjects and adds new information as follows
$${c_t} = {f_t} \times {c_{t - 1}} + {\tilde c_t} \times {i_t}.$$

Output gate decide what we’re going to output and reading from Memory which is formulated as
\ba
{o_t} = \sigma ({W_o} \times [{h_{t - 1}},{x_t}] + {b_o})\nonumber \\
{h_t} = {o_t} \times \tanh ({c_t}).
\ea
\\

\subsection{Data and Features}

In this paper, we use four major (open, high, low, and closing prices) Oil stock pairings-  FP.PA (Total), CNE.L (Cairn Energy), BP.L (BP), SLB.PA (Schlumberger) with their daily interval data and use for comparison stock pairings- WTI, gold, dollar with their daily interval data from 2009/8/14 to Sep 2020/7/19 for \textbf{training} and data from 2020/7/20 to 2021/7/15 for \textbf{testing}.Data was extracted from Alpha Vantage API.\cite{alpha}
\\

This section provides a summary of the findings obtained through experiments conducted using our proposed methodology. The neural network model utilized was the LSTM network, implemented using the Python Neural Networks library and Keras, running within the TensorFlow 2.0 Python development environment to train the data.
\subsection{Hyper-parameter selection and Network topology}
\begin{figure}[!htb]
	\centering
	\includegraphics[width=0.8 \columnwidth]{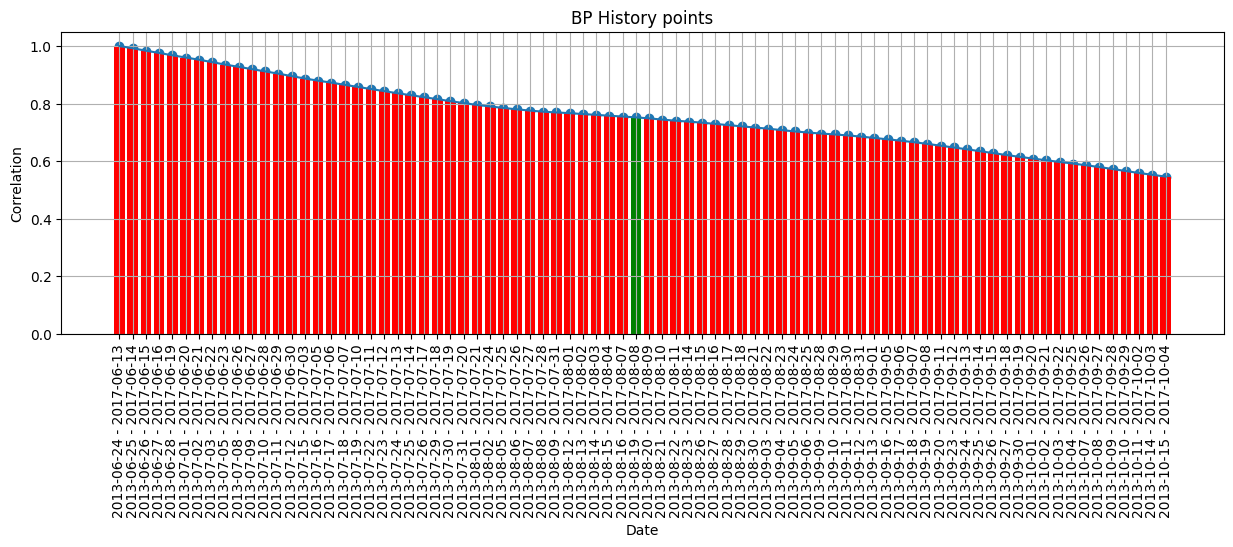}
	\caption{BP past and future correlations  .}
	\label{bp-hp}
\end{figure}

\begin{figure}[!htb]
	\centering
	\includegraphics[width=0.8 \columnwidth]{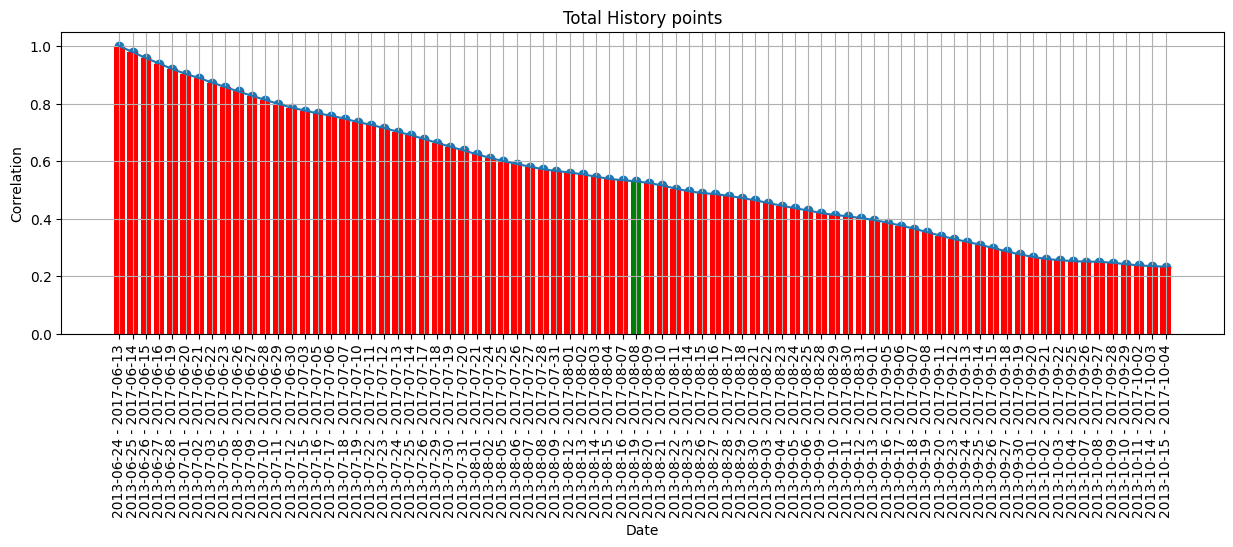}
	\caption{Total past and future correlations .}
	\label{total-hp}
\end{figure}

\begin{figure}[!htb]
	\centering
	\includegraphics[width=0.8 \columnwidth]{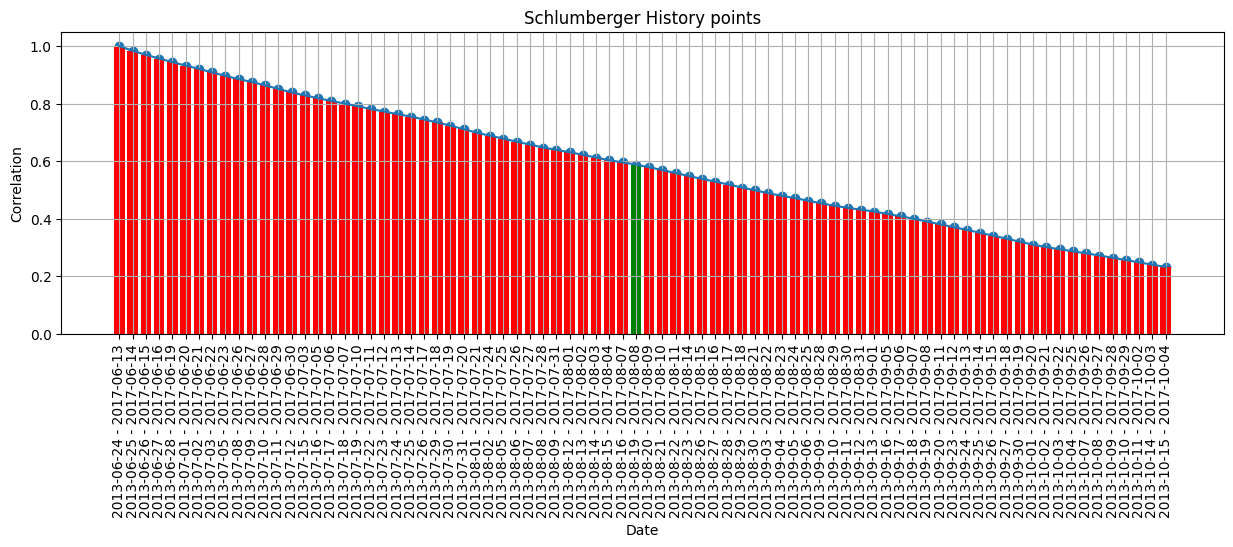}
	\caption{SLB past and future correlations .}
	\label{slb-hp}
\end{figure}

\begin{figure}[!htb]
	\centering
	\includegraphics[width=0.8 \columnwidth]{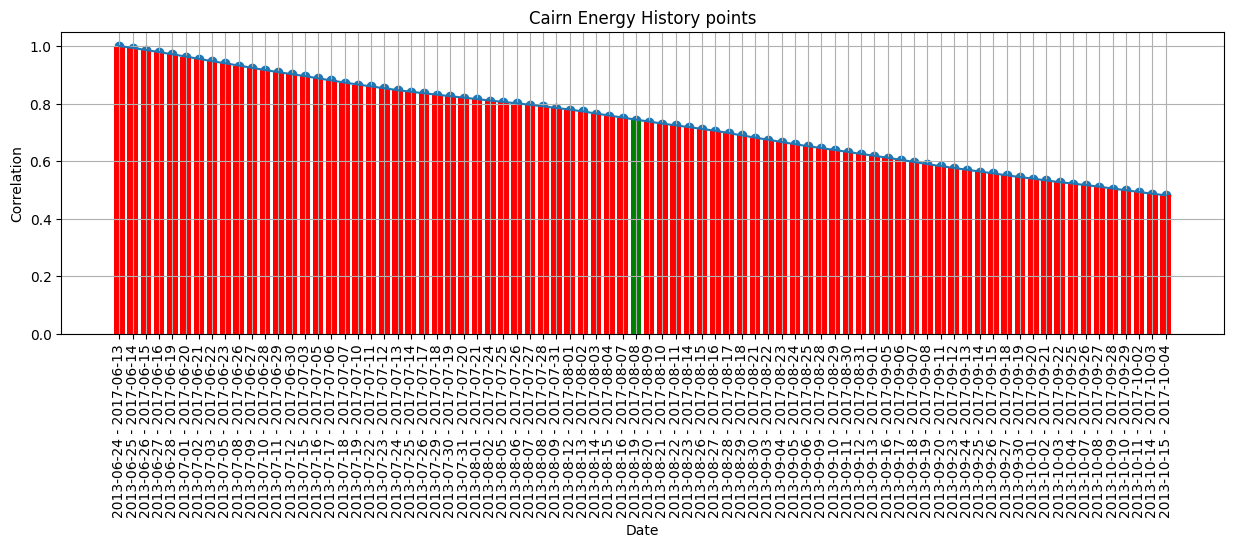}
	\caption{Cairn Energy past and future correlations .}
	\label{cne-hp}
\end{figure}

The fundamental determination of hyper-parameters for the neural network has a very important effect on the modeling results.appropriate number of time lags is one of the most important hyper-parameters that has a very important effect on the results in modeling with recurrent neural networks.Autocorrelation is a statistical method used to measure the relationship between a time series data point and its lagged values. It indicates whether there is a correlation between data points that are a certain distance apart in time. Autocorrelation can be used in the context of Long Short-Term Memory (LSTM) models to find the appropriate number of time lags to use for prediction.In LSTM models, the previous time steps are used as features for predicting the next time step. One of the key challenges is to determine the appropriate number of time lags to use for prediction. If too few lags are used, important features may be missed, leading to poor performance. If too many lags are used, the performance may also suffer due to overfitting.Autocorrelation can be used to address this challenge by measuring the correlation between the time series data point and various lagged values. By plotting the autocorrelation function, we can determine the optimal number of lags to use for prediction. In general, the autocorrelation function declines as the lag increases, with the most significant lags occurring at the beginning of the time series.Basic determination of this value prevents.The results of the Autocorrelations of the close price of each stock from 2013/6/23 to 2017/6/12 with each of the dates in Fig. \eqref{bp-hp}, Fig. \eqref{total-hp}, Fig. \eqref{slb-hp} and Fig. \eqref{cne-hp}are as follows:\\
\\
\\
\\
\\
Due to the fact that in the \eqref{dcorr} section, we have discretized the data as 40-day packages and then performed the Autocorrelation analysis, and according to Fig. \eqref{bp-hp}, \eqref{total-hp}, \eqref{slb-hp}, and \eqref{cne-hp}, we can see that the data has a strong direct relationship with the future 40 days.According to the obtained results, we set the value of the appropriate number of time lags equal to 40, so that we can check the interpretability of LSTM for predicting the data of oil stocks based on the results obtained from the modeling and the results of the \eqref{dcorr} section.\\
Adam's\cite{adam} optimization function is used with 0.0005 learning rate, 10 percent data is validation data, epochs value is equal to 50 and batch size is equal to 25.For all simulations with LSTM, these values are the same.\\

For modeling, we have used a simple standard LSTM topology, which is displayed in Table\eqref{tp-table}.\\

\begin{table}[!htb]
	\caption{This table illustrates the network topology which includes each layer along with neurons and computational parameters.}
	\label{tp-table}
	\begin{tabular} {l|l|l}
		\textbf{Layers}    & \textbf{Neurons} &	\textbf{Computational parameters}\\ \hline
		Inputs             & (None,50,6)   &	0	  \\
		LSTM               & (None,50)     &  11400   \\
		Dropout            & (None,50)     &	0     \\
		Dense              & (None,64)     &	3264  \\
		Activation(Sigmoid)& (None,64)     &  0       \\
		Dense              & (None,1)      &	65    \\
		Activation(Linear) & (None,1)      &  0       \\
		&&\\
		\textbf{Total parameters}&14729&\\
		\textbf{Trainable parameters}&14729&\\
		\textbf{Non-trainable parameters}&0&\\
	\end{tabular}
\end{table}
\subsection{Evaluation Measures}

To measure the badness of the model, people use Mean Square Error (MSE), Root means square error (RMSE), Mean Absolute Error (MAE), and  Mean Absolute Percentage Error (MAPE), as defined below:
For measuring the average of the squares of the errors, MSE measure can be used which is defined as the following formula:
$$MSE = {{\sum\nolimits_1^n {{{(true - Prediction)}^2}} } \over n}.$$
\\
To measure the average root of the squares of the errors we use RMSE which is defined as:
$$RMSE = \sqrt {{{\sum\nolimits_1^n {{{(true - Prediction)}^2}} } \over n}}, $$
\\
which describes how spread out these residuals are by the standard deviation of prediction errors. To measure the average of absolute error we use MAE which calculates errors between pair observations. It is defined as:
$$MAE = {{\sum\nolimits_1^n {|true - Prediction|} } \over n}.$$
\\
Finally, to measure the size of the error in percentage terms we use MAPE which is calculated for the mean of the absolute percentage errors of prediction. It is defined as the following formula:
\be
MAPE = {{{{\sum\nolimits_1^n {|true - Prediction|} } \over {true}}} \over n} \times 100.
\ee
Now, we are in the position where we can apply these evaluation measures as error metrics in our prediction.\\

\section{Discussion}
In this study, we have demonstrated the potential of deep learning as a tool for forecasting stock prices in the oil industry, either as a replacement or a complement to traditional forecasting methods. 

We began by investigating the correlation coefficients between the stocks of various oil companies, as well as the USD, WTI crude oil, and gold indices. We then discretized the stock data and performed discrete correlation analysis.

In Sec.\eqref{corr}, we observed that there is a correlation between the stocks of the oil companies in question and the gold, dollar, and crude oil indices. Based on the findings presented in Table \eqref{corr-table}, we obtained the following results: \\

Overall, our analysis suggests that there are significant correlations and relationships among the stock indices of the oil companies, as well as with the USD, WTI, and gold indices. Specifically, BP and Cairn Energy stock indices show strong direct relationships with each other, while TOTAL appears to be more inversely correlated with gold. Schlumberger has weaker correlations with the other indices, but exhibits a strong direct relationship with TOTAL and weak direct relationships with BP and Cairn Energy. Our findings suggest that deep learning algorithms may be particularly useful in predicting stock prices within the oil industry given these complex interrelationships and correlations.

Secondly, in order to verify the accuracy and stability of the results obtained from Section \eqref{corr}, and prior to modeling and checking the results of the LSTM model, we will examine the results of Section \eqref{dcorr} which presents a discrete analysis of the correlations.


Secondly, in order to verify the accuracy and stability of the results obtained from Section \eqref{corr}, and prior to modeling and checking the results of the LSTM model, we will examine the results of Section \eqref{dcorr} which presents a discrete analysis of the correlations.
\begin{itemize}
\item According to Fig.\eqref{bpcorr}, BP shares display fluctuations in their correlations with the index of gold, USD, and WTI stocks throughout abnormal correlation intervals, which are not confined to specific correlation intervals. In Fig. \eqref{bp-hist} and Table \eqref{bp_corr-table}, around 30 percent of the correlations exhibit weak direct relationships with WTI, which contradicts the results of Table \eqref{corr-table}, whereas approximately 40 percent show a strong direct association. About 36 percent of the correlations indicate an inverse relationship with the USD, which is contrary to Table \eqref{corr-table}, but around 64 percent of the correlations demonstrate a strong direct relationship with the USD. Approximately 46 percent of the correlations suggest an inverse relationship with gold, while about 54 percent of the correlations demonstrate a direct relationship with gold, which contradicts the findings of Table \eqref{corr-table}. The average correlation of BP shares with the WTI stock index is 0.27, indicating a weak direct association, which is consistent with the results of Table \eqref{corr-table}. However, this average has a standard deviation of 0.45, representing high dispersion. The median value of 0.32 is noteworthy, acting as the intermediary between the outcome obtained from Table \eqref{corr-table} and the average of the correlations, thereby indicating that the correlations with an inverse relationship carry more weight. This may suggest the influence of indirect factors such as political events. The average correlation of BP's shares with the gold stock index almost indicates a lack of relationship and negative correlation, which contradicts the findings of Table \eqref{corr-table}. The median value is smaller than the average, indicating that the correlations outweigh the inverse relationship between the two stocks. The average correlation of BP shares with the dollar suggests a weak direct relationship, contradicting the results of Table \eqref{corr-table}. Moreover, the average of these correlations highlights the weight of the correlations with a direct relationship.
\item Similar to BP's correlations shown in Fig. \eqref{bpcorr}, shares of TOTAL, Schlumberger, and CNE companies display high fluctuations in the correlation ranges based on Fig.\eqref{totalcorr}, Fig. \eqref{slbcorr} and Fig. \eqref{cnecorr}. According to Fig. \eqref{total-hist} and Table \eqref{total_corr-table}, roughly 60 percent of the correlations demonstrate a direct relationship between TOTAL shares and the stock index and WTI, which contradicts the results of Table \eqref{corr-table}. About 56 percent of the correlations exhibit a direct relationship between TOTAL shares and the USD, which contradicts the findings of Table \eqref{corr-table}, but in terms of gold, it confirms the results of Table \eqref{corr-table}. The median correlations of TOTAL shares with the USD and gold are approximately zero, which aligns with the results of Table \eqref{corr-table} in the case of USD shares, but in the case of gold, it contradicts the outcomes of Table \eqref{corr-table}. This is an essential point. Based on the standard deviation value, there is a high dispersion among the correlation values. The average and median correlations of this stock with WTI are consistent with the results of Table \eqref{corr-table}.
\item Based on Fig. \eqref{slb-hist} and Table \eqref{slb_corr-table}, approximately 70 percent of the correlations imply a direct relationship between Schlumberger company shares and WTI shares, matching the results of Table \eqref{corr-table}. About 56 percent of the correlations suggest a direct relationship between Schlumberger shares and the USD, while roughly 52 percent of the correlations contradict the outcomes of Table \eqref{corr-table} regarding their relationship with gold stocks. The mean correlations in Table \eqref{cne-static-t} for Schlumberger stocks with WTI show a weak direct relationship. The mean and average of these shares with gold stocks are both zero, which contradicts the results of Table \eqref{corr-table}. Furthermore, concerning the mean and median of these shares with the USD, it indicates a greater weight of correlations that demonstrate a direct relationship.
\item Based on Fig. \eqref{cne-hist} and Table \eqref{cne_corr-table}, around 72 percent of the correlations indicate a direct relationship between Cairn Energy shares and WTI, consistent with the outcomes of Table \eqref{corr-table}. About 70 percent of the correlations demonstrate a direct relationship between these stocks and the USD, contradicting the results of Table \eqref{corr-table}, which suggested an absence of a relationship. Roughly 52 percent of the correlations suggest a direct relationship between Cairn Energy shares and gold shares, contradicting the outcomes of Table \eqref{corr-table}. The median and average correlations of Cairn Energy shares with the USD imply a weak direct relationship, differing from the results of Table \eqref{corr-table}. The mean and average of these shares with gold show a contradiction with the outcome of Table \eqref{corr-table} by indicating a lack of a relationship. Also, the mean and average correlations of these stocks with WTI demonstrate a direct relationship, aligning with the results of Table \eqref{corr-table}.
\end{itemize}
Based on the results obtained from the discrete analysis of correlations, we can observe that the correlation outcomes are not always consistent and do not necessarily increase within a specific correlation interval. Instead, they fluctuate across different correlation intervals. In the oil markets, various factors like political decisions, shifts in politics, and conflicts impact the price, which changes over time.
Third, to forecast stock prices of different companies, we employed Recurrent Neural Networks with LSTM architecture, given that these stocks exhibit changes in time series. We conducted empirical experiments and performed on the stock indices dataset to evaluate the predictive performance in terms of several common error metrics, namely MSE, MAE, RMSE, and MAPE. Here are the summary results based on these metrics:
\begin{itemize}
	\item In Table \eqref{bp-table}, it was observed that including the WTI (crude oil), gold, and dollar indices did not enhance the model's prediction and did not lessen its errors (as seen in the error metrics). Furthermore, the same four primary features had the lowest cost function. The empirical analysis results for BP are illustrated in Fig. \eqref{bp-fig}:
\begin{table}[!htb]
	\caption{Experiment Result For BP Shares}
	\label{bp-table}
	\begin{tabular} {l|l|l|l|l}
		& MSE&	RMSE&	MAE&	MAPE  \\ \hline
		 WTI&	0.02634&	0.16232	&0.01026&	0.08321\\
		Main Dataset&	0.00372&0.06105&0.00386&0.08335\\
		 Dollar&	0.13510&	0.36756&	0.02324&	0.08096\\
		 Gold &	0.02252&	0.15008&	0.00949&	0.08311\\
	\end{tabular}
\end{table}
\begin{figure}[!htb]
	\centering
	\includegraphics[width=1 \columnwidth]{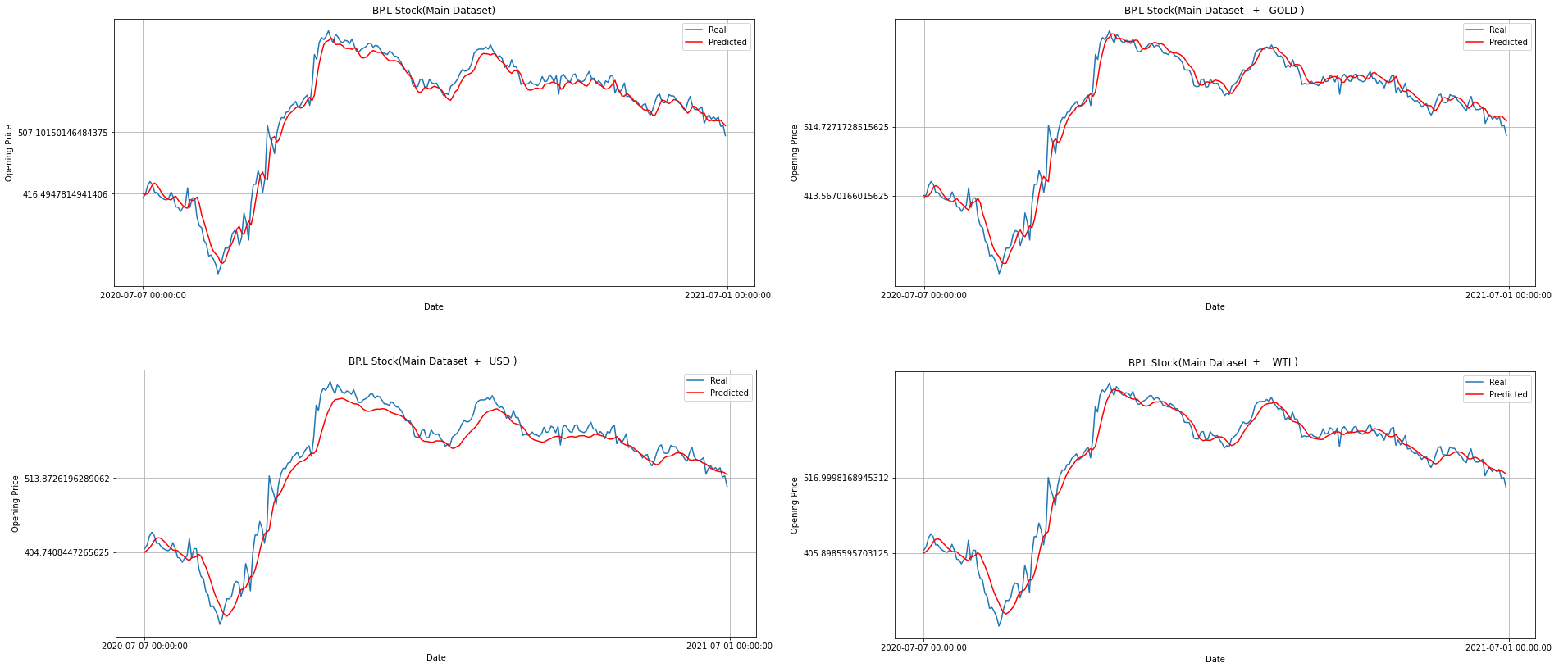}
	\caption{(BP Real Prices vs Predicted Prices) }
	\label{bp-fig}
\end{figure}

\item In Table \eqref{cne-table}, it was observed that incorporating the crude oil and dollar indices did not decrease the cost function. However, during the learning process, when the machine analyzed the gold index, it significantly improved the measurement and resulted in reduced costs, thereby helping us better our modeling, as depicted in the Cairn Energy diagrams in Fig. \eqref{cne-fig}.

\begin{table}[!htb]
	\caption{Experiment Result For Carien Energy Shares}
	\label{cne-table}
	\begin{tabular} {l|l|l|l|l}
		& MSE&	RMSE&	MAE&	MAPE  \\ \hline
		 WTI&	146.4406	&12.1012	&0.7653	&0.3544\\
		Main Dataset&	78.5752&	8.8642&	0.5606&	0.4298\\
		 Dollar &79.4293&	8.9123&	0.5636&	0.4272\\
		 Gold&	48.4487&	6.9605&	0.4402&	0.4733\\
	\end{tabular}
\end{table}
\begin{figure}[!htb]
	\centering
	\includegraphics[width=1 \columnwidth]{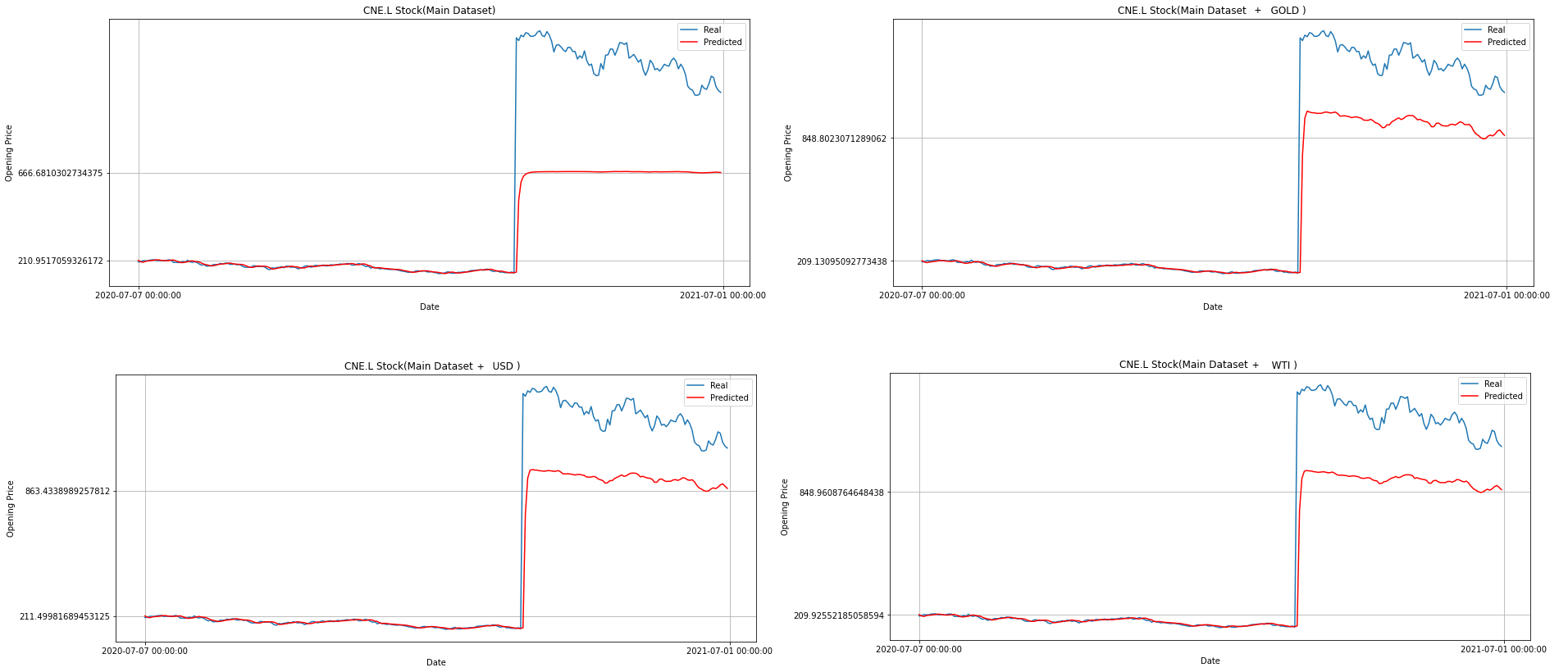}
	\caption{(Cairn Energy Real Prices vs Predicted Prices) }
	\label{cne-fig}
\end{figure}

\item Table \eqref{sch-table} indicates that the WTI index resulted in an increase in the cost function, and the addition of the gold and dollar indices made no significant impact on the learning process. The diagrams for the Schlumberger company are illustrated in Fig. \eqref{sch-fig}.
\begin{table}[!htb]
	\caption{Experiment Result For Schlumberger Shares}
	\label{sch-table}
	\begin{tabular} {l|l|l|l|l}
		& MSE&	RMSE&	MAE&	MAPE  \\ \hline
		 WTI&0.00964&	0.09818&	0.00621&	0.02830\\
		Main Dataset&0.00102&	0.03203&	0.00202&	0.02569\\
		 Dollar&0.00112&	0.03361&	0.03361&	0.025234\\
		 Gold&0.00108&	0.03296&	0.00208&	0.025690\\
	\end{tabular}
\end{table}
\begin{figure}[!htb]
	\centering
	\includegraphics[width=1 \columnwidth]{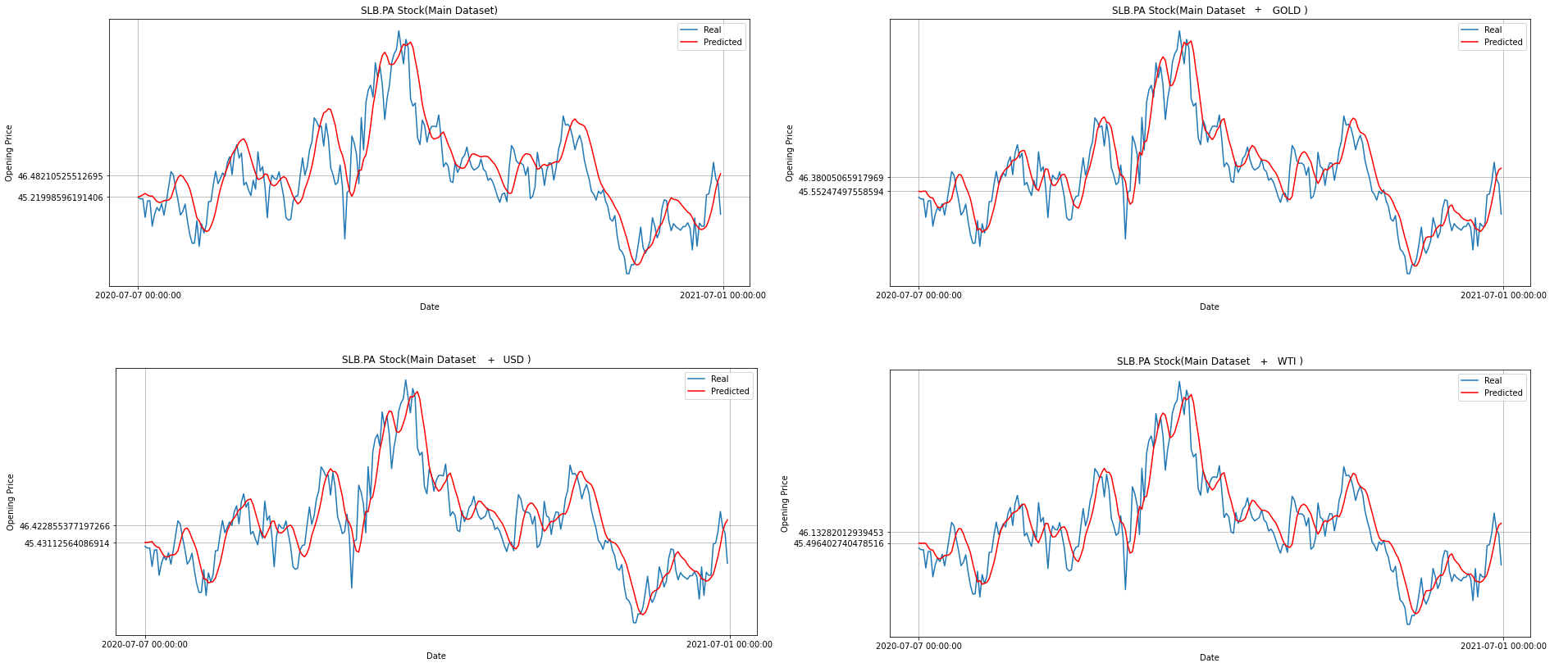}
	\caption{(Schlumberger Real Prices vs Predicted Prices) }
	\label{sch-fig}
\end{figure}

\item Table \eqref{fp-table} regarding Total Shares reveals that none of the WTI, dollar, or gold indicators enhanced the learning process, as evidenced by Fig. \eqref{fp-fig}.
\begin{table}[!htb]
	\caption{Experiment Result For Total Shares}
	\label{fp-table}
	\begin{tabular} {l|l|l|l|l}
		& MSE&	RMSE&	MAE&	MAPE  \\ \hline
		 WTI&	0.000297&	0.01723	&0.00109&	0.02621\\
		Main Dataset&	2.9648e-07&	0.0005442&	3.4437e-05&	0.02582 \\
		 Doler&	2.3122e-05&	0.00480	&0.00030&	0.02613\\
		 Gold	&0.00040	&0.02001	&0.00126&	0.02579\\
		
	\end{tabular}
\end{table}
\begin{figure}[!htb]
	\centering
	\includegraphics[width=1 \columnwidth]{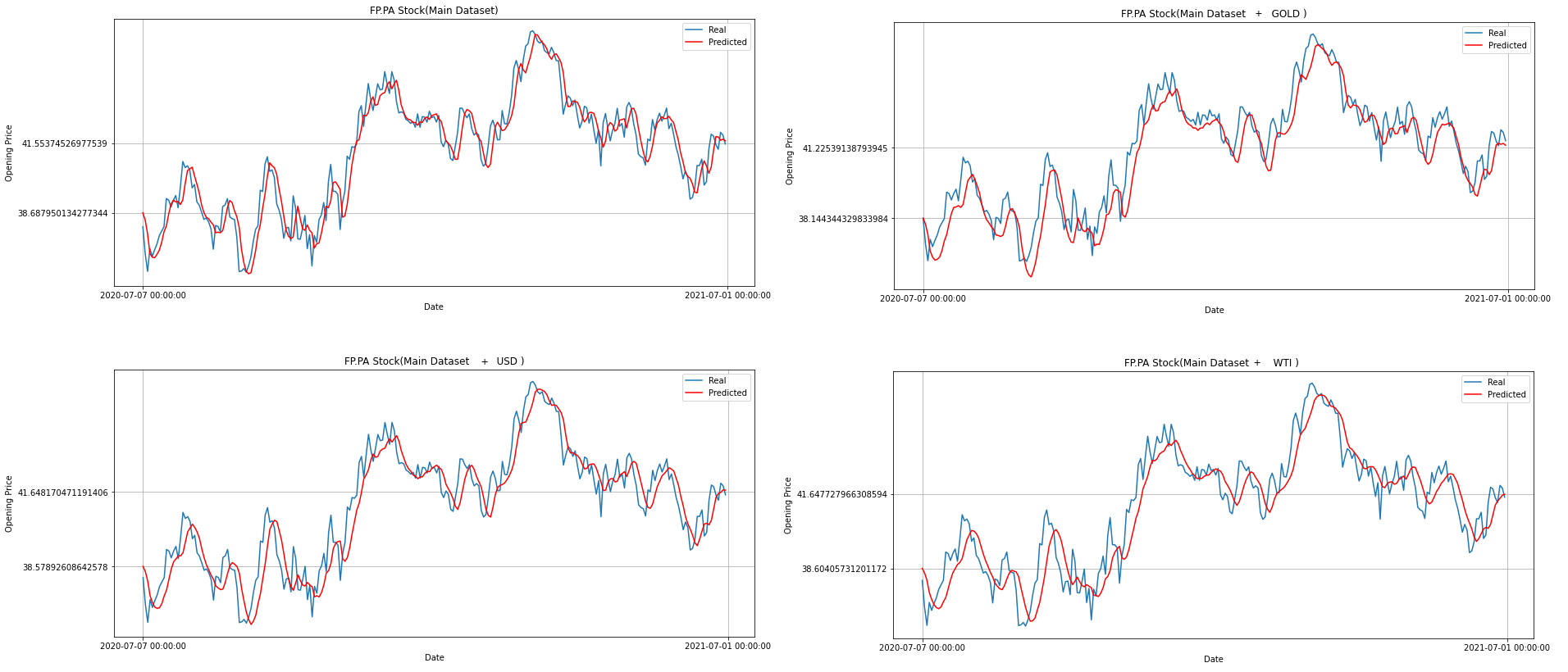}
	\caption{(Total Real Prices vs Predicted Prices) }
	\label{fp-fig}
\end{figure}
\end{itemize}

\section{Conclusion}
In general, the presented results show that the range of errors for each stock varies among different models, though these differences are consistent across all stocks. Based on our analysis, it can be concluded that the use of correlated features did not improve the performance of the LSTM model for the oil market. 

Our findings revealed that although LSTM models could achieve high accuracy in predicting stock prices, their interpretability was limited. Specifically, their internal states and weight parameters were difficult to interpret and could not provide clear insights into the underlying factors driving stock price movements. Overall, caution should be exercised when relying solely on LSTM models for stock price prediction, as their lack of interpretability may make it difficult to fully understand the factors driving stock price movements. Further research is needed to improve the interpretability of LSTM models for predicting stock prices.\\

{\bf Acknowledgments:}\\

We would like to thank M. H. Zhoolideh Haghighi for useful comments and discussion.\\


\end{document}